\documentclass[preprint2]{aastex}

\usepackage{lineno}
\usepackage{amsmath}
\usepackage{graphicx}
\usepackage{natbib}
\usepackage{hyperref}

\newcommand{\pvec}{\boldsymbol{p}}
\newcommand{\xvec}{\boldsymbol{x}}

\renewcommand{\vec}[1]{\boldsymbol{#1}}
\newcommand{\braket}[1]{\langle #1\rangle}

\newcommand{\del}{\partial}

\newcommand{\ed}[1]{\textbf{#1}}
\renewcommand{\ed}[1]{#1}

\shorttitle{Ionization in atmospheres of Brown Dwarfs and extrasolar planets IV}
\shortauthors{Rimmer \& Helling}

\begin{document}


\title{Ionization in atmospheres of Brown Dwarfs and extrasolar planets IV. The Effect of Cosmic Rays}


\author{P. B. Rimmer and Ch. Helling}
\affil{SUPA, School of Physics \& Astronomy, University of St Andrews, St Andrews, KY16 9SS, UK}
\email{pr33@st-andrews.ac.uk}



\begin{abstract}
Cosmic rays provide an important source for free electrons in the Earth's atmosphere and also in dense interstellar regions where they produce a prevailing background ionization. We utilize a Monte Carlo 
cosmic ray transport model for particle energies of $10^6$ eV $< E < 10^9$ eV, and an analytic cosmic ray transport model for particle energies of $10^9$ eV $ < E < 10^{12}$ eV in order to investigate 
the cosmic ray enhancement of free electrons in substellar atmospheres of free-floating objects. The cosmic ray calculations are applied to \textsc{Drift-Phoenix} model atmospheres of an example brown dwarf 
with effective temperature $T_{\rm eff} = 1500$ K, and two example giant gas planets ($T_{\rm eff} = 1000$ K, $1500$ K). For the model brown dwarf atmosphere, the electron fraction is enhanced significantly 
by cosmic rays when the pressure $p_{\rm gas} < 10^{-2}$ bar. Our example giant gas planet atmosphere suggests that the cosmic ray enhancement extends to $10^{-4} - 10^{-2}$ bar, depending 
on the effective temperature. For the model atmosphere of the example giant gas planet considered here ($T_{\rm eff} = 1000$ K), cosmic rays bring the degree of ionization to 
$f_e \gtrsim 10^{-8}$ when $p_{\rm gas} < 10^{-8}$ bar, 
suggesting that this part of the atmosphere may behave as a weakly ionized plasma. Although cosmic rays enhance the degree of ionization by over three orders of magnitude in the upper atmosphere, the effect 
is not likely to be significant enough for sustained coupling of the magnetic field to the gas.
\end{abstract}





\section{Introduction}
\label{sec:intro}

M-dwarfs of spectral class M7 and hotter have been discovered to emit
quiescent X-rays, and lower mass brown dwarfs emit X-rays
intermittently \citep{Berger2010}. Currently, there is no satisfactory
explanation for this phenomenon, but there are some promising models
that may account for the observed X-ray emissions. For example,
\citet{Helling2011,Helling2011b} propose that charge build-up on
grains may lead to atmospheric ionization to a degree sufficient to
couple the magnetic field to a partially ionized atmospheric gas. The
magnetic fields would follow the convective atmospheric dynamics, and
would become tangled. X-rays would then be the consequence of magnetic
reconnection events. This scenario requires a charge density at least
$10^6$ times greater than predicted by thermal ionization in current
model atmospheres \citep[see][their Fig. 2]{Helling2011}. Other
ionizing mechanisms need therefore to be considered. In
  this paper, we provide a first study of how significant cosmic ray
  ionization is in the atmospheres of brown dwarfs and giant gas
  planets.  

If cosmic rays contribute significantly to atmospheric
ionization in extrasolar planets and brown dwarfs, the effects of this
ionization may provide an opportunity to better constrain the
energy spectrum of galactic cosmic rays in a variety of planetary atmospheres. 
A large number of cosmic rays are blocked from Earth by the solar wind \citep{Jokipii1976},
although the Voyager probe is expected to measures the interstellar cosmic ray flux \citep{Webber2009}. 
Free-floating planets and stellar objects would not be so protected, and the observable 
effects of cosmic ray ionization in the atmospheres of these objects may provide us an
indirect way to determine the spectrum of extrasolar cosmic rays.

Cosmic rays were first discovered because of their ionizing effect on
the earth's atmosphere \citep{Hess1912}.  Cosmic ray ionization may
affect terrestrial climate conditions by enhancing aerosol formation
\citep{Pudovkin1995,Shumilov1996} and initiating discharge events
\citep{Ehrmakov2003,Stozhkov2003}.
\citet{Griessmeier2005} investigate the possible impact of cosmic ray showers on biological organisms
in extrasolar earth-like planets with weak magnetic fields by considering 
the effect of a weaker planetary magnetic field on cosmic ray propagation through the planetary atmosphere.
 The effect of the Earth's electric field on cosmic ray
propagation is also being explored. For example, \cite{Muraki2004}
found measurable enhancement in cosmic ray intensity when a negative
electric field of magnitude $>10$ kV/m is present in the
atmosphere. \cite{Bazilevskaya2008} provide a comprehensive review of
the research into connections between cosmic rays and the
atmosphere. The effect of cosmic rays on the exospheres of free-floating
  extrasolar planets and brown dwarfs that form dust clouds in the low atmosphere 
has not been explored so far.
The impact of cosmic rays on the exosphere is of particular interest
as it links the underlaying atmosphere with the object's
  magnetosphere and may also help to understand coronal effects in the
  substellar mass regime.  

The cosmic ray opacity of
brown dwarf and hot jupiter atmospheres has been explored by
\cite{Helling2012}, where the cosmic ray flux is considered to
decrease exponentially with the column density of the gas
\citep{Umebayashi1981}. The effect of cosmic ray propagation on the rate of
ionization of the gas has been modeled in some detail for diffuse conditions in
the interstellar medium \citep{Padovani2009,Rimmer2012}, dense
environments \citep{Umebayashi1981} and the terrestrial atmosphere
\citep{Velinov2009}. \citet{Rimmer2012} and \citet{Velinov2009} both
employ Monte Carlo models for cosmic ray propagation. \citet{Velinov2009}
consider pair-production and particle decay \ed{ (full Monte-Carlo 
simulation of an atmospheric cascade)}, and \citet{Rimmer2012} consider the 
energy loss due to ionizing collisions and the effect of a weak magnetic 
field ($B < 1$ mG) on the cosmic ray spectrum.
\ed{ The models of \citet{Velinov2008,Velinov2009} have been tested against 
various atmospheric profiles and model assumptions \citep{Mishev2008,Mishev2010}.}
 \citet{Padovani2009} numerically 
solve the propagation integral from \\
\citet{Cravens1978}, and account for energy 
lost due to ionization and excitation. \citet{Umebayashi1981} solve a set of 
Boltzmann transport equations for the cosmic rays, with a term for 
energy loss due to pair-production of electrons.

In this paper, we explore the impact of cosmic ray ionization on the
electron fraction in model atmospheres of an example brown dwarf and two example
giant gas planets. 
To this end, we utilise model atmospheres which do not 
necessarily resample any known free-floating planets.
In order to separate the effect of cosmic rays
from external UV photons, we only consider \emph{free-floating} objects
not in proximity to a strong UV field.  We utilize 1D Monte
  Carlo and analytic cosmic ray propagation methods over a wide range of densities
to allow a principle investigation of the effect of cosmic rays.

In order to explore cosmic ray ionization in the atmosphere,
  propagation of the cosmic rays through both the exosphere and atmosphere must be
  treated. A simple density profile for the gas in the exosphere is
  calculated in Section \ref{sec:exosphere}, using the Boltzmann
  Transport Equation and assuming a Maxwellian distribution for the
  gas. This is necessary because the exosphere is very likely no longer
a  collisionally dominated gas, hence the continuum assumption for 
applying hydrodynamic concepts breaks down \citep[see, e.g.][]{Chamberlain1987}.  

We are using {\sc Drift-Phoenix} model atmosphere structures which are the
result of the solution of the coupled equations of radiative
transfer, convective energy transport (modelled by mixing length
theory), chemical equilibrium (modelled by laws of mass action),
hydrostatic equilibrium, and dust cloud formation (Dehn 2007, Helling
et al. 2007a,b; Witte, Helling \& Hauschildt 2009, \citealt{Helling2011}). The dust cloud
formation model includes a model for seed formation (nucleation),
surface growth and evaporation of mixed materials, and gravitational
settling (drift) (Woitke \& Helling 2003, 2004; Helling \& Woitke
2006; and Helling, Woitke \& Thi 2008). The results of the {\sc
Drift-Phoenix} model atmosphere simulations include the
gas temperature - gas pressure structure ($T_{\rm gas}$, $p_{\rm
gas}$), the local gas-phase composition in thermochemical equilibrium, 
the local electron number
densities ($n_{\rm e}$), and the number of dust grains ($n_{\rm d}$)
and their sizes ($a$) dependent on the height in the atmosphere. These
models are determined by the effective temperature, $T_{\rm eff}$, the
surface gravity, log(g), and the initial elemental abundances. 
The elemental abundances are set to the solar values throughout 
this paper.

We consider a model atmosphere of a brown dwarf, with a surface gravity of
$\log g = 5$ and an effective temperature of $T_{\rm eff} = 1500$ K, as well
as model atmospheres for two example giant gas planets, both with $\log g = 3$.
The effective temperatures of free-floating exoplanets are not constrained, allowing
us to freely explore this part of the parameter space. We chose the values of $T_{\rm eff} = 1500$ K
to allow for direct comparison with earlier work \citep{Helling2011}, and $T_{\rm eff} = 1000$ K,
which is the inferred effective temperature of \object{HR 8799 c}, a giant gas planet $\sim 40$ AU
from its host star \citep{Marois2008}. 
The distance of $40$ AU is far enough to possibly allow the planet's parameters 
to be effectively the same as those of a free-floating giant gas planet.

 Section \ref{sec:crt} describes our cosmic
ray transport calculations for cosmic rays of energy $10^6$ eV $< E <
10^{12}$ eV. We then calculate the steady-state degree of ionization by cosmic rays. 
We combine the cosmic ray ionization to the thermal degree of ionization from \textsc{Drift-Phoenix}. 
Section \ref{sec:results} contains the resulting degree of ionization and includes the effect on the 
coupling of the magnetic field of the gas.

\section{Density Profile of the Exosphere}
\label{sec:exosphere}

\begin{figure}
\centering
\begin{tabular}{c}
\includegraphics[width=0.8\columnwidth]{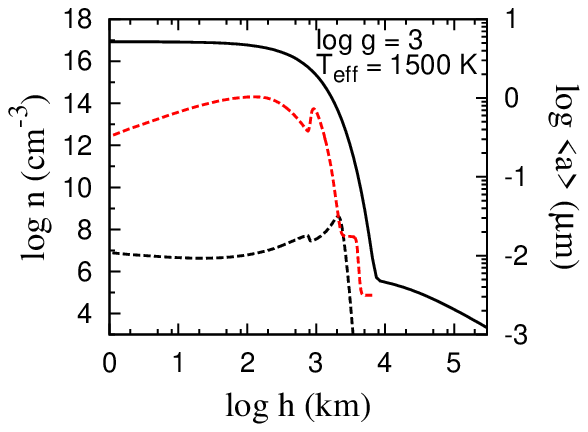} \\
\includegraphics[width=0.8\columnwidth]{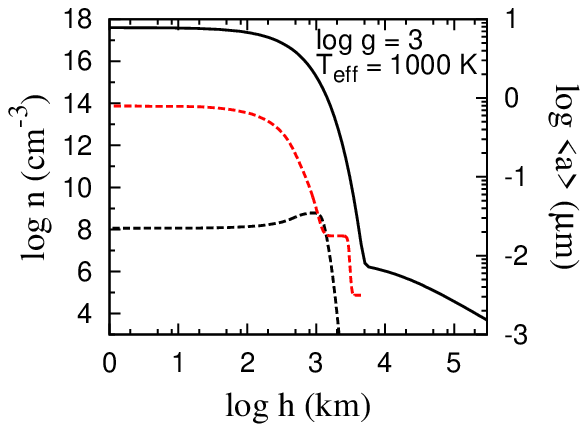}\\
\includegraphics[width=0.8\columnwidth]{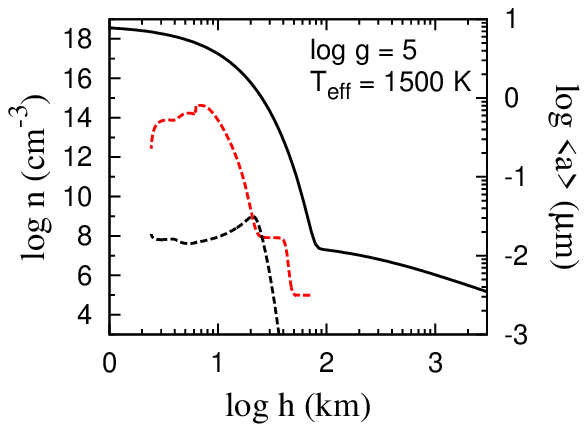}
\end{tabular}
\caption{The total number density of the exospheric and atmospheric gas, $n_{\rm gas}$ [cm$^{-3}$] 
(solid black line, left axis) and of the cloud particles, $n_{\rm dust}$ [cm$^{-3}$] (dashed black line, 
left axis), and the mean grain radius, $\braket{a}$ [$\mu$m] (dashed red line, right axis), versus 
atmospheric height, $h$ (km), for $\log g = 3$, $T_{\rm eff} = 1500$ K (top), $\log g = 3$, 
$T{\rm eff} = 1000$ K (middle) and $\log g = 5$, $T_{\rm eff} = 1500$ K (bottom).
The plots show that the cloud layers extend to considerably lower pressures in giant gas planets than in
brown dwarfs.}
\label{fig:profile}
\end{figure}

For modelling purposes, we divide the gas around the planet or
brown dwarf into three regions (inward $\rightarrow$ outward): the cloud layer\footnote{The 
cloud layer is part of the atmosphere, because the cloud particles form from the atmospheric gas.
Figure \ref{fig:profile} indicates the location and extent of the cloud layer.}
(lowest), the dust-free upper atmosphere (middle) and the exosphere (highest).  
The location and extent of the cloud layer is determined by the dust formation and atmosphere model
\textsc{Drift-Phoenix} \citep{Dehn2007,Helling2008,Witte2009}. The
exosphere is considered here to be the regime where the gas can no
longer be accurately modeled as a fluid. This occurs when the mean
free path of the gas particles is of the order of the atmospheric
scale height.  The pressure at which this is the case is the lowest
pressure considered in the \textsc{Drift-Phoenix} model atmospheres,
and we place the exobase at that height.

\begin{figure}
\centering
\begin{tabular}{c}
\includegraphics[width=0.5\columnwidth,height=0.6\columnwidth]{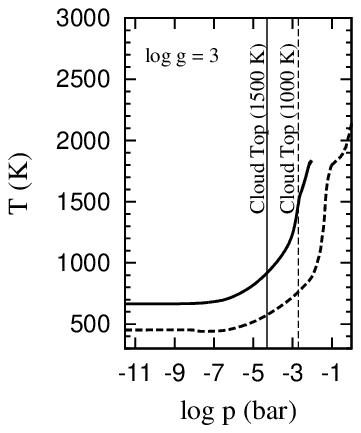}
\includegraphics[width=0.4\columnwidth,height=0.6\columnwidth]{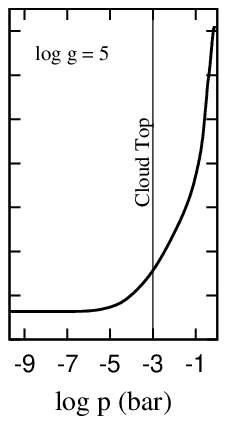}
\end{tabular}
\caption{The pressure-temperature profiles for model atmospheres of three example planets, with the parameters
 $\log g = 3$, $T_{\rm eff} = 1500$ K (left, solid), $\log g = 3$, $T_{\rm eff} = 1000$ K and 
$\log g = 5$, $T_{\rm eff} = 1500$ K (right). The vertical 
line on each plot indicates the cloud top. Pressures less than the cloud
top pressure are considered to be within the upper atmosphere.}
\label{fig:pT}
\end{figure}

Figure \ref{fig:profile} provides gas density and cloud particle
 number density profiles of the three example model atmospheres and exospheres considered here, as well as
 the mean grain sizes of cloud particles for the model atmosphere of two example giant gas planets 
($\log g = 3$, $T_{\rm eff} = 1000$ K, $1500$ K) and a brown dwarf ($\log g = 5$, $T_{\rm eff} = 1500$ K).
This grain size profile demonstrates the location of the 
cloud with respect to the atmospheric temperature and gas density. 
Figure \ref{fig:pT} shows the $(p,T)$ profiles for the
model atmospheres.

In order to model cosmic ray transport into a planet's atmosphere, it
is necessary to treat all material between the atmosphere and the
source of the incident (galactic) cosmic ray spectrum.  Since galactic
cosmic ray propagation models \citep{Strong1998} and observations of
chemical tracers \citep{Indriolo2007} currently suggest that cosmic
rays of energies $10^6$ eV $ \lesssim E \lesssim 10^{12}$ eV are
ubiquitous in the diffuse interstellar medium, we treat the cosmic ray
spectrum to be that of the galactic spectrum at the ``upper
edge''\footnote{Technically, there is no definitive upper edge to the
  exosphere. The gas density decreases monotonically, but there is no
  definitive transition past the exobase. The location of an ``upper
  edge'' is therefore somewhat arbitrary. We treat the ``upper edge'' 
of the exosphere to be an infinite distance from the model planet or brown dwarf.} 
of the exosphere, and initiate cosmic ray transport at that point.

We calculate the density profile of the exosphere by solving the steady-state collisionless 
Boltzmann Transport Equation with a gravitational force term:
\begin{equation}
 \dfrac{\pvec}{m}\cdot\nabla f + m\vec{g} \cdot \dfrac{\del f}{\del \pvec} = 0,
 \label{eqn:BTE} 
\end{equation}
where $\pvec$ is the momentum vector, $m$ is the particle mass,
$\vec{g} = GM/r^2 \vec{\hat{r}}$ is the gravitational field at radial displacement
$\vec{r}$ from the center of mass, $G = 6.67 \times 10^{-8}$ cm$^{3}$
g$^{-1}$ s$^{-2}$, and $M$ is the mass of the giant gas planet or
brown dwarf.  $f(\xvec,\pvec)$ [cm$^{-6}$ g$^{-3}$ s$^{3}$] is the
distribution function, representing the number of particles in the
volume-element $d^3x \, d^3p$ at location ($\xvec,\pvec$) in the phase
space.  We treat the exosphere one-dimensionally for the sake of
simplicity. Denoting $(r,p_r)$ as the coordinates of interest, $g =
GM/r^2$, and Eq. (\ref{eqn:BTE}) becomes:
\begin{equation}
 \dfrac{p_r}{m} \, \dfrac{df(r,p_r)}{dr} - \dfrac{GmM}{r^2} \dfrac{df(r,p_r)}{dp_r} = 0.
 \label{eqn:BTE-1D}
\end{equation}
Now $f(r,p_r)$[cm$^{-4}$ g$^{-1}$ s] is a one dimensional distribution
function representing the number of particles located within a volume element $d^3x$ at $r$, with 
momentum between $p_r$ and $p_r + dp_r$. Both Maxwell-Boltzmann and Lorentzian distribution functions
have been applied to the exosphere \citep{Pierrard1996}.  The
Maxwell-Boltzmann distribution has been applied to interstellar
conditions of similar temperature and density to exosphere conditions,
and the calculated deviations from this distribution in the
interstellar environment is generally found to be on the order of
$1\%$ \citep[][his Sect. 2.3]{Spitzer1978}. We therefore choose the
Maxwell Boltzmann distribution function for our exosphere model.  The
Maxwell Boltzmann distribution is:
\begin{equation}
 f(r,p_r) = \dfrac{n_{\rm gas}}{(2\pi m k_B T)^{1/2}} e^{-p_r^2/2mk_BT}, 
 \label{eqn:distribution}
\end{equation}
where $T$ is the temperature of the gas, $k_B = 1.4 \times 10^{-16}$ erg/K 
denotes the Boltzmann constant, and $n_{\rm gas}$ [cm$^{-3}$] is
the gas density profile. We apply Eq. (\ref{eqn:distribution}) to 
Eq. (\ref{eqn:BTE-1D}). We note that $f$ includes $n_{\rm gas} = n_{\rm gas}(r)$ and $T = T(r)$.
Employing the product rule for differentiation, the first term of 
Eq. (\ref{eqn:BTE-1D}) can be written as:
\begin{align}
  \dfrac{p_r}{m} \dfrac{df(r,p_r)}{dr} &= \dfrac{p_r}{m}f(r,p_r)\dfrac{1}{n_{\rm gas}}\dfrac{dn_{\rm gas}}{dr} \notag\\
& - \dfrac{p_r}{2m}f(r,p_r)\dfrac{1}{T}\dfrac{dT}{dr} \notag\\
& + \dfrac{p_r^3}{2m^2k_B}f(r,p_r)\dfrac{1}{T^2}\dfrac{dT}{dr},
 \label{eqn:one}
\end{align}
and the second term is:
\begin{equation}
 -\dfrac{GmM}{r^2} \dfrac{df(r,p_r)}{dp_r} = \dfrac{p_r}{k_BT}\dfrac{GM}{r^2}f(r,p_r).
 \label{eqn:two}
\end{equation}
We now reconstruct the transport equation:
\begin{align}
  \dfrac{p_r}{m}f(r,p_r)\dfrac{1}{n_{\rm gas}}\dfrac{dn_{\rm gas}}{dr} - \dfrac{p_r}{2m}f(r,p_r)\dfrac{1}{T}\dfrac{dT}{dr} & \notag\\
+ \dfrac{p_r^3}{2m^2k_B}f(r,p_r)\dfrac{1}{T^2}\dfrac{dT}{dr} + \dfrac{p_r}{k_BT}\dfrac{GM}{r^2}f(r,p_r) &\notag\\ 
= 0 & ,
  \label{eqn:new-BTE}
\end{align}
We multiply by $m/p_r$ and integrate this equation over $p_r$. The first two terms in Eq. (\ref{eqn:new-BTE}) become:
\begin{align}
 \int_{-\infty}^{\infty} f(r,p_r)\dfrac{1}{n_{\rm gas}}\dfrac{dn_{\rm gas}}{dr}\;dp_r &= \dfrac{dn_{\rm gas}}{dr};\\
 \int_{-\infty}^{\infty} \dfrac{1}{2}f(r,p_r)\dfrac{1}{T}\dfrac{dT}{dr}\;dp_r &= \dfrac{n_{\rm gas}}{2}\dfrac{1}{T}\dfrac{dT}{dr};
\label{eqn:terms-1-2}
\end{align}
The third term in Eq. (\ref{eqn:new-BTE}) becomes:
\begin{align}
 & \int_{-\infty}^{\infty} \dfrac{p_r^2}{2mk_B} f(r,p_r)\dfrac{1}{T^2}\dfrac{dT}{dr}\;dp_r \notag\\ 
 & = \dfrac{n_{\rm gas}}{\sqrt{\pi}(2mk_BT)^{3/2}} \, \dfrac{1}{T}\dfrac{dT}{dr} \int_{-\infty}^{\infty} p_r^2 \, e^{-p_r^2/2mk_BT} \; dp_r, \notag\\
 & = \dfrac{n_{\rm gas}}{2}\dfrac{1}{T}\dfrac{dT}{dr}.
\label{eqn:term-3}
\end{align}
 and the last term becomes:
 \begin{equation}
  \int_{-\infty}^{\infty} \dfrac{1}{k_BT}\dfrac{GmM}{r^2}f(r,p_r)\;dp_r = \dfrac{GmM}{r^2}\dfrac{n_{\rm gas}}{k_BT}
  \label{eqn:term-4}
 \end{equation}
 Applying Eqn's (\ref{eqn:terms-1-2})-(\ref{eqn:term-4}) to the integral of Eq. (\ref{eqn:new-BTE}) 
over $p_r$, the Boltzmann Transport Equation becomes (with explicit $r$-dependence):
\begin{equation}
 \dfrac{1}{n_{\rm gas}(r)}\dfrac{dn_{\rm gas}(r)}{dr} = -\dfrac{GmM}{k_B r^2T(r)}
 \label{eqn:BTE-Maxwellian}
\end{equation}
This equation is equivalent to the equation for the distribution function for quasi-collisionless
exospheres from \citet[][their Eq. 7.1.16]{Chamberlain1987}, with the angular momentum set 
to zero and temperature as a function of the radial distance. 
In order to solve this equation, the temperature would have to be determined 
 via radiative transfer. If we
were instead to treat the temperature as a constant in
the R.H.S. of Equation (\ref{eqn:BTE-Maxwellian}), the equation
would become identical to the condition for hydrostatic equilibrium,
but this condition is not appropriate for the exosphere. The next
simplest functional form for the temperature is the result of setting
the average thermal energy, $(3/2) \, k_B T$, equal to the Virial of
the system, which assumes that the system as a whole is stable and
bounded. The temperature then becomes:
\begin{equation}
 T(r) = \dfrac{GmM}{3k_Br}.
 \label{eqn:virial-temp}
\end{equation}
Applying Equation (\ref{eqn:virial-temp}) to Eq. (\ref{eqn:BTE-Maxwellian}), we find:
\begin{equation}
 \dfrac{1}{n_{\rm gas}(r)}\dfrac{dn_{\rm gas}(r)}{dr} = - \dfrac{3}{r},
 \label{eqn:virial-diff}
\end{equation}
with the solution:
\begin{equation}
 n_{\rm gas}(r) = n_c \Bigg(\dfrac{r_c}{r}\Bigg)^3.
 \label{eqn:diff-simple}
\end{equation}
The inner boundary condition, at the height of the exobase (denoted as $r_c$) is that the number density $n_{\rm gas}(r_c)$
is the number density at the exobase, $n_c$.

\begin{figure}
\centering
\includegraphics[width=\columnwidth]{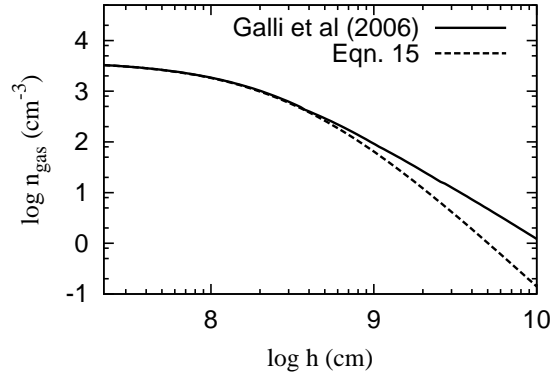}
\caption{Total gas number density of the Martian exosphere as a function of height 
above the surface, $h = r - R_{\rm Mars}$, where $R_{\rm Mars} = 3400$ km is
the radius of Mars. The solid line represents an analytical fit to the exospheric 
model predictions of \cite[][their Eq. 4]{Galli2006} and
the dashed line represents our analytical calculations for the Martian exosphere density, 
from Eq. (\ref{eqn:diff-simple}). Our ansatz for the exosphere density profile 
underestimates the Galli exospheric density profile by about an order of magnitude at $h = 10^{10}$ cm.}
\label{fig:exosphere}
\end{figure}

For a sanity check, we compare our results to the more detailed models for the Martian exosphere \citep{Galli2006}. 
\citet{Galli2006} provide an analytic form of the density profile for the Martian exosphere,
and compare it to observation. We are not aware of a more recent analytic expression for exospheric
densities that is compared to observational results.
Taking their value for the exobase, $r_c = 3.62 \times 10^8$ cm (220 km above the surface of Mars), setting our
$r = h + R_M$, where $R_M = 3.40 \times 10^8$ cm is the radius of Mars,
and using the density at the exobase from their Eq. (4) for our $n_c$, we can compare their
model results with our own Figure \ref{fig:exosphere}. The profile of \citet[][their Eq. 4]{Galli2006}
is within an order of magnitude to our profile for $r < 10^{10}$ cm. Since the two profiles differ significantly
only at great distances, when $n \ll n_c$, the impact these differences have on the column density is rather small, namely within
a factor of $2$ as $r \rightarrow \infty$.

By the use of the Virial theorem, we have neglected any thermal emission from the atmosphere. The
correct treatment requires a radiative transfer simulation of the atmosphere,
which is beyond the scope of this paper.
We do not expect these uncertainties to significantly effect cosmic ray transport on free-floating
planets and brown dwarfs, for reasons detailed in Section \ref{sec:crt}. Other relevant uncertainties arise from the
neglected planet's rotation, the gravitational pull
and radiation pressure from the sun. Heating from external sources of
radiation can also impact the density profile
\citep{Hinteregger1981,Watson1981,Lammer2003,Vidal2004,Murray2009}.
Since we are neglecting external sources of UV
radiation throughout this paper, the differences arising from
radiation pressure and heating will not concern us here. If external
UV radiation were to be incorporated (i.e. for young and/or active stars), the impact on exospheric
properties could be quite significant. \citet{Murray2009} found the effect of external
photons, not considered in this study, on the exospheric density profile of up to two orders of magnitude.

Since we consider the exosphere to start at the outermost point in the
atmosphere for the \textsc{Drift-Phoenix} model atmosphere under
consideration, $n_0$ (Eq.~\ref{eqn:diff-simple}) is taken to be the
outermost density from that model.  For the sample brown dwarf atmosphere we
consider here ($\log(g) = 5$, $T_{\rm eff} = 1500$ K, solar
metallicity), the exobase density, $n_c = 3 \times 10^9$ cm$^{-3}$
which is at $\sim 100$ km above the cloud layer.  For the sample giant gas planet
($\log(g) = 3$, $T_{\rm eff} = 1500$ K, solar metallicity), $n_c = 4
\times 10^7$ cm$^{-3}$ which is at $\sim 5000$ km above the cloud
layer. 

To determine the column density of the exosphere, $N_{exo}$
[cm$^{-2}$], through which cosmic rays must travel before reaching the
upper atmosphere, we perform the integral:
\begin{equation}
 N_{exo} = \int_{r_c}^{\infty} n_{\rm gas}(r) \; dr,
 \label{eqn:colden}
\end{equation}
and applying Equation (\ref{eqn:diff-simple}) to Equation
(\ref{eqn:colden}), we have that $N_{exo} \approx \frac{1}{2} \, n_c r_c$.
Assuming the radii of both the $\log g = 3$ and $\log g = 5$ cases to
be the radius of Jupiter, $R_J = 7.1 \times 10^9$ cm, $N_{exo} \approx
5 \times 10^{18}$ cm$^{-2}$ for brown dwarfs and $\approx 7.5 \times 10^{16}$
cm$^{-2}$ for a giant gas planet \citep[e.g.][]{Chabrier2000,Burrows2001}.
Although our model is one-dimensional and neglects
rotation and gravitational interaction with
other bodies, the cosmic ray transport is not very sensitive to the
rather low exospheric column density. Column densities $\lesssim 5 \times 10^{19}$ cm$^{-2}$
will not significantly affect cosmic ray transport, according to our model, 
explained below in Section \ref{sec:crt}. Errors in the exospheric gas
density profile by up to an order of magnitude do not significantly
effect our results for cosmic ray transport.

\section{Cosmic Ray Transport in the Exosphere and Atmosphere}
\label{sec:crt}

We have now determined the amount of material a cosmic ray will have
to pass through before reaching a certain depth of an
atmosphere. In order to determine the average number of cosmic rays to
reach a given atmospheric pressure, we must now consider how much
energy cosmic rays lose through ionizing collisions, and how many
cosmic rays are lost through electron-positron production, pion decay, muon
decay, and several other processes.
This will allow us to determine both how far into the
atmosphere cosmic rays of a given energy reach, and their electron
production once they are there. Solving the cosmic ray transport
  i.e. the collisional interaction of the individual cosmic ray
  particles with the surrounding gas, will allow us to estimate an
ionization rate for cosmic rays as a function of penetration depth
into the atmosphere. The exospheric column density, $N_{exo}$, will be
applied to cosmic ray transport along with the column density between
the exobase and a given depth into the atmosphere, $N_{atm}$, and
$N_{col} = N_{exo} + N_{atm}$. In this section, we will determine a
cosmic ray ionization rate that depends on the total column,
$N_{col}$.

Cosmic rays of kinetic energy, $E$ [eV], are divided here into
low-energy cosmic rays (LECRs, $E < 10^9$ eV) and high-energy cosmic
rays (HECRs, $E > 10^9$ eV). LECR transport was calculated by
\cite{Rimmer2012} using a Monte-Carlo transport model that
incorporates the magnetic field effects detailed in
\citet{Skilling1976} and \citet{Cesarsky1978}, as well as
inelastic collision energy loss. Collisional energy loss is included
by first taking the average energy loss per collision, $W(E)$ [eV],
from \cite{Dalgarno1999}, and the ``optical'' depth for the cosmic ray. 
This depth is $\sigma(E) \, \Delta N$, 
where $\sigma(E)$ is the total cross-section for inelastic collisions 
between protons. We use the cross-sections from
\cite{Rudd1983} and \cite{Padovani2009}.

\begin{figure}
\centering
\begin{tabular}{cc}
\includegraphics[width=0.47\columnwidth,height=0.5\columnwidth]{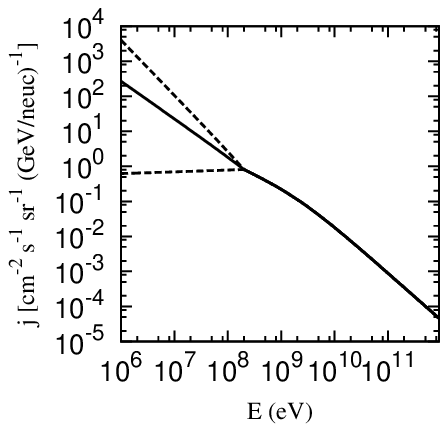} &
\includegraphics[width=0.4\columnwidth,height=0.5\columnwidth]{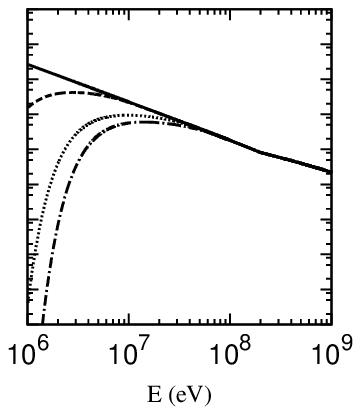}
\end{tabular}
\caption{The flux spectrum of cosmic rays, $j(E)$ [cm$^{-2}$ s$^{-1}$
    sr$^{-1}$ (GeV/nucleon)$^{-1}$] versus the cosmic ray energy, $E$,
  from Eq. (\ref{eqn:init-flux}, left), and as it varies with column density, $N_{\rm col}$,
according to our Monte Carlo cosmic ray propagation model (right). The left plot shows the effect of varying the parameter $\alpha$ 
that describes the power-law component of the spectrum
below $2 \times 10^8$ eV.
The solid line is for $\alpha =
  -2.15$, the dashed lines bound $-3.15 < \alpha < 0.1$.
The right plot shows the flux-spectrum with $\alpha = -2.15$ at $N_{\rm col} = 0$ cm$^{-2}$ (solid),
$1.5 \times 10^{21}$ cm$^{-2}$ (dashed), $10^{22}$ cm$^{-2}$ (dotted) and $2.5 \times 10^{22}$ cm$^{-2}$ (dash-dot).
}
\label{fig:spectrum}
\end{figure}

We apply a Monte Carlo model to determine LECR transport. In this model,
detailed in \citet{Rimmer2012}, we take 10000 cosmic rays, and assign
each of them two numbers. The first number corresponds to the energy of the
individual cosmic ray, and the energies are distributed among the cosmic
rays according to the initial cosmic ray flux spectrum (see Fig. \ref{fig:spectrum}). The
second number is a random number of uniform distribution with range $[0,1]$. 
The cosmic rays are advanced through the atmosphere over a column, 
$\Delta N$. If the random number is less
than $\sigma(E) \, \Delta N$, the cosmic ray collides with an atmospheric particle,
and loses an amount of energy, $W(E)$. We choose the size $\Delta N$ so that $\sigma_{\rm max} \, \Delta N < 1$,
where $\sigma_{\rm max}$ is the maximum cross-section for an inelastic collision between a proton and a hydrogen atom, 
$\approx 5 \times 10^{-16}$ cm$^2$ \citep[see][their Eqn's. 5,10]{Padovani2009}.
A new spectrum is generated by binning the cosmic rays according to their energies, and then the cosmic rays are advanced
another segment of the column, and the process is repeated.

At the very end of the process, we have a series of cosmic ray spectra, from the initial
galactic cosmic ray spectrum at the edge of the exosphere, to the cosmic ray spectrum at the
bottom of the cloud layer. The strength of Alfv\'en waves generated by
cosmic rays depends on the difference between the initial cosmic ray
spectrum and a given spectrum within the atmosphere.

\citet{Rimmer2012} consider the bulk of galactic cosmic rays to be positively charged (protons and alpha particles). 
This is the typical assumption, and is currently supported by observation \citep{Webber1998}. These cosmic rays lose energy in the
exosphere and atmosphere through ionizing collisions, until they are eventually thermalized. The result is a charge imbalance, with more positive charges
present higher in the exosphere than in the exobase or upper atmosphere. This charge imbalance causes electrons to move from the exobase and upper atmosphere
higher into the exosphere, in order to neutralize the positive charge. These electrons will attempt to drag the magnetic field lines with them, generating
Alfv\'{e}n waves. The cosmic rays therefore lose energy in proportion to the energy of the Alfv\'{e}n waves generated, in addition to the energy lost from collisions
with the ambient gas. This mechanism for energy
loss was first examined by \citet{Skilling1976}, who considered cosmic ray exclusion in the interstellar medium, where magnetic fields are on the order of $3 \; \mu$G.
If the magnetic field is much stronger than this ($\gtrsim 1$ mG), then the electrons will be more strongly locked to the magnetic field lines and they will no longer
be able to efficiently generate Alfv\'{e}n waves (i.e. the inequality in Eq. 4 in \citealt{Rimmer2012} will no longer be satisfied). Since this mechanism accelerates
free electrons within the atmosphere, this may allow a cosmic ray driven current in atmospheric regions where 
the local magnetic field strength is $ \leq 1$ mG.

The question of the effect strong magnetic fields have on cosmic ray propagation
is beyond the scope of this principle investigation. Large scale magnetic fields for brown dwarfs are several orders
of magnitude greater than our 1 mG limit \citep{Reiners2007}, so a study of strong magnetic field effects on cosmic ray transport
is of great importance. \citet{Griessmeier2005} investigated cosmic ray propagation in exoplanet exospheres and atmospheres
in the presence of earth-strength magnetic fields, and found a correlation between the strength of the magnetic
moment and anisotropy of cosmic rays on the surface of the planet \citep[][their Fig. 5]{Griessmeier2005}.
This anisotropy effect will increase for brown dwarfs as the strength of their magnetic field is suggested to be larger.

We take an initial flux density of cosmic rays, $j(E)$ [cm$^{-2}$
  s$^{-1}$ sr$^{-1}$ ($10^9$ eV/nucleon)$^{-1}$] to be the
broken power-law spectrum from \cite{Indriolo2009}, based on the the models of \\
\cite{Lerche1982} and \cite{Shibata2006}. This spectrum best fits
both the observed light element abundances produced via scintillation
as well as observed abundances of the ion H$_3^+$ in the interstellar
medium. The broken power-law spectrum changes sharply in flux at $< 2
\times 10^8$ eV. This change results from ``leaky-box'' models 
\citep{Lerche1982,Shibata2006}, and is argued to be caused by
low-energy shocks from either supernova remnants or possibly OB stellar
atmospheres interacting with the ambient medium
\citep{Bykov1992}. The initial flux density used in this paper
 is given as \citep[based on][their Eq. 8]{Indriolo2009}:
\begin{equation}
 j =
\begin{cases}
  j(E_1) \bigg( \dfrac{p(E)}{p(E_1)} \bigg)^{\gamma}, & \text{if } E > E_2 \\
  j(E_1) \bigg( \dfrac{p(E_2)}{p(E_1)} \bigg)^{\!\!\gamma} \!\! \bigg( \dfrac{p(E)}{p(E_2)} \bigg)^{\!\!\alpha}, & \text{if } E_{\rm cut} < E < E_2 \\
  0, & \text{if } E < E_{\rm cut}
\end{cases}
\label{eqn:init-flux}
\end{equation}

where $p(E) = \frac{1}{c}\sqrt{E^2 + 2EE_0}$ and $E_0 = 9.38 \times
10^8$ eV is the proton rest-energy. $E_1 = 10^9$ eV and $E_2 = 2
\times 10^{8}$ eV are constants, and the flux $j(E_1) = 0.22$
cm$^{-2}$ s$^{-1}$ sr$^{-1}$ [$10^9$ eV/nucleon]$^{-1}$ is the
measured cosmic ray flux at $10^9$ eV. The fitting parameter $\gamma
\approx -1.35$ is also well-constrained by observation \citep{Mori1997}. 
The value for the second fitting exponent
$\alpha$ depends on the models of \cite{Lerche1982} and
\cite{Shibata2006}; we choose the value $\alpha = -2.15$ 
because this value best agrees with the cosmic ray ionization 
in the interstellar medium inferred by \citet{Indriolo2007}.
The flux-spectrum in Eq. (\ref{eqn:init-flux})  above $E \sim 10^6$ eV is 
expected to be roughly the same throughout the interstellar medium \citep{Strong1998},
and therefore seems to be a sensible initial cosmic ray flux applied at the
outer edge of our exosphere.
The parameter $\alpha$ determines the hardness
of the LECR component of the spectrum, and has no observed lower
limit, because LECR's are shielded from us by the solar wind. Upper
limits to $\alpha$ can be determined from Voyager observations
\citep[e.g.][]{Webber1998}. The effect on the spectrum from varying $\alpha$
can be seen in Figure \ref{fig:spectrum}.
This figure also includes a plot showing how the cosmic ray spectrum 
changes with column-density into our model atmospheres. An application of the
results of \cite{Rimmer2012} to cosmic rays below $10^6$ eV shows that
cosmic rays are unlikely to travel farther than $\sim 1$ parsec from their
source of origin, the origin being either a supernova remnant or an OB
association.  It is therefore sensible to apply a low-energy cut-off,
$E_{\rm cut} = 10^6$ eV, to the initial cosmic ray spectrum applied
in this paper.

It is important to determine how the ionization rate changes 
when traveling into the exosphere and into the atmosphere. Fewer cosmic
rays will be able to reach deeper into the atmosphere, 
so cosmic rays will affect the ambient gas less and less with increasing atmospheric
depth.  
We now determine the cosmic ray
ionization rate as a function of column-density.  We apply
Eq. (\ref{eqn:init-flux}) to the Monte Carlo model from
\cite{Rimmer2012} to determine a column-dependent flux-density. This
flux-density, $j(E)$, can be used to calculate to the primary cosmic ray ionization rate
for Hydrogen, $\zeta_p$ [s$^{-1}$] by:
\begin{equation}
 \zeta_p = 4 \pi (1 + G_{10}) \int_0^{\infty} \big[1+\phi_p(E)\big] j(E) \, \sigma_p(E) \; dE.
 \label{eqn:primary-zeta}
\end{equation}
Here, $\sigma_p(E)$ is the ionization cross-section for a proton
to ionize a hydrogen atom \citep{Spitzer1968,Padovani2009} and
$G_{10}$ is a factor representing the ionization by LECR electrons and
by ``heavy'' LECRs (mostly $\alpha$-particles), and is assumed to be
$G_{10} = 0.8$ \citep{Spitzer1968}. The ionizing event produces a free electron at
super-thermal energies, which often causes the ionization of additional
species. This is accounted for in the term $\phi_p(E)$ which takes the form from
\citet{Glassgold1973,Padovani2009} of:
\begin{equation}
 \phi_p(E) = \dfrac{1}{\sigma_p(E)}\int_{I({\rm H_2})}^{\infty} P(E,E_{\rm se}) 
\, \sigma_{\rm se}(E_{\rm se}) \; dE_{\rm se},
\label{eqn:secondary-phi}
\end{equation}
where $E_{\rm se}$ is the energy of the secondary electron, $I({\rm H_2}) = 15.603$ eV is the
ionization energy of molecular hydrogen, $\sigma_e$ [cm$^2$]
is the cross-section for ionization of H$_2$ by an electron \citep{Mott1930} and
$P(E,E_{\rm se})$ is the probability of the secondary electron having energy $E_{\rm se}$
given a proton of energy $E$. We approximate the Eq. (\ref{eqn:secondary-phi}) as:
\begin{equation}
  \phi_p(E) \approx \dfrac{\sigma_{\rm se}\big(W(E) - I({\rm H_2})\big)}{\sigma_p(E)},
\label{eqn:phi-approx}
\end{equation}
where $W(E)$ is the average amount of energy deposited into the secondary electron from the cosmic ray
proton, from \citet{Dalgarno1999}. If $W(E) < I({\rm H_2})$ then $\phi_p = 0$.

\begin{figure}
\centering
\includegraphics[width=0.8\columnwidth]{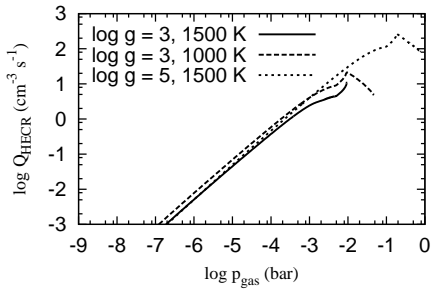}
\caption{The electron production rate from cosmic rays 
of energies $E > 10^9$ eV ($Q_{\rm HECR}$ [cm$^{-3}$ s$^{-1}$]), versus pressure, $p_{\rm gas}$ [bar], for 
both model giant gas planet atmospheres, with $\log g = 3$ ($T_{\rm eff} = 1500$ K (solid line) and 
$T_{\rm eff} = 1000$ K (dashed line)) and brown dwarf atmosphere with $\log g = 5$, $T_{\rm eff} = 1500$ K
(dashed line). The electron production rate, $Q_{\rm HECR}$, is obtained analytically from \citet{Velinov2008}.
For the $\log g = 3$ (giant gas planet) cases, the electron production peaks at 
$10^{-2}$ bar (1000 K) and $0.25$ bar (1500 K).
Electron production for the $\log g = 5, 1500$ K (brown dwarf) case does not have a clear peak over the range of
pressures we considered.}
\label{fig:e-production}
\end{figure}

For HECRs, electrons are primarily produced by electron-positron
  production, muon decays and other high-energy effects, and less so by
  ionizing collisions for which the cross-section is very small at
  high energies. \cite{Velinov2009} apply the CORSIKA \ed{code\footnote{\ed{The CORSIKA code incorporates several models for hadron-hadron 
interactions as well as various parameterizations of the Earth's atmosphere.}}} to
 \ed{realistic} terrestrial atmospheric conditions \ed{\citep[see][]{Mishev2008,Mishev2010}} 
in order to calculate the rate of electron production from HECRs, and find that it compares well with
their analytical method \citep[see also][]{Velinov2008}. The high-energy cosmic ray
penetration is significantly affected by atmospheric composition \citep{Molina1999}. We therefore adapt 
the analytical method from \citet{Velinov2008} to our atmosphere by
making the following changes. Where \citet{Velinov2008} set the ionization energy to that of 
nitrogen, $I({\rm N_2})$, for the terrestrial atmosphere, 
we use an averaged ionization energy for the our atmospheric chemistry,
\begin{equation}
I_{\rm Av} \approx \dfrac{I({\rm H_2})n({\rm H_2}) + I({\rm He})n({\rm He}) + I({\rm H})n(\rm{H})}{n_{\rm gas}}.
\label{eqn:ionization-energy}
\end{equation}
We apply the same methods to determine an average proton number, $Z_{\rm Av}$, 
and average mass number, $A_{\rm Av}$, for our atmospheric
models. We also set the $E_{\rm min} = 10^9$ eV \citet[from][their Eq. 25,27]{Velinov2008}, 
because we treat low-energy cosmic ray
transport separately. The results of the analytic model of \citet{Velinov2008}, in 
terms of the electron production by high energy cosmic rays, $Q_{\rm HECR}(N_{col})$ [cm$^{-3}$ s$^{-1}$],
for giant gas planets and brown dwarf atmospheres, are shown in Figure \ref{fig:e-production}.
The production rate of electrons depends on the gas column density 
according to \citet[][their Eq. 26,28]{Velinov2008}. The electron production for our 
$\log g = 3$ model is within an order of magnitude of the electron production calculated 
for Jupiter's atmosphere from \citet[][their Fig. 3a]{Whitten2008}.

We combine the HECR results of \cite{Velinov2008,Velinov2009} to the
the Monte Carlo calculations for LECRs from \cite{Rimmer2012}. This simply amounts to taking a total cosmic ray ionization rate, $Q$ [cm$^{-3}$ s$^{-1}$]:
\begin{equation}
 Q(N_{col}) = n_{\rm gas} \zeta_{\rm LECR}(N_{col}) + Q_{\rm HECR}(N_{col}).
\end{equation}
 The column-dependent ionization rate as a function of column density, $N_{col}$ [cm$^{-2}$], 
is well fit by the analytical form:
\begin{align}
 Q & (N_{col}) =  Q_{\rm HECR}(N_{col}) \notag\\ 
& +\zeta_0 n_{\rm gas} \times \begin{cases}480 & \!\!\!\!\! \text{if } N_{col} < N_1 \\
   1 + (N_0/N_{col})^{1.2} & \!\!\!\!\!  \text{if } N_1 < N_{col} < N_2\\
  e^{-kN_{col}} & \!\!\!\!\!  \text{if } N_{col} < N_2\\
 \end{cases}
 \label{eqn:e-production}
\end{align}
where $\zeta_0 = 10^{-17}$ s$^{-1}$ is the standard ionization rate 
in the dense interstellar medium, and the column densities $N_0 = 7.85 \times 10^{21}$ 
cm$^{-2}$, $N_1 = 4.6 \times 10^{19}$ cm$^{-2}$, 
$N_2 = 5.0 \times 10^{23}$ cm$^{-2}$, and $k = 1.7 \times 10^{-26}$ cm$^2$ 
are fitting parameters. Our calculated brown dwarf exospheric column density is within a 
factor of 3 of $N_1$, and the giant gas planet exospheric 
column density is about two orders of magnitude below $N_1$. In both cases, 
we do not expect the exosphere to have much impact on our cosmic ray 
transport model for giant gas planets and brown dwarfs. For atmospheric depths
above $\approx 1.7$ g cm$^{-2}$, Eq. (\ref{eqn:e-production}) converges to 
\citet[][their Eq. 1]{Helling2012}.

We calculate the degree of ionization using the same method as \citet{Whitten2008}. We can estimate the steady-state number density of electrons, $n(e^-)$ [cm$^{-3}$], by the rate equation:
\begin{equation}
\dfrac{dn(e^-)}{dt} = Q - \alpha_{\rm DR} \big[n(e^-)\big]^2,
\label{eqn:ionization-rate-eqn}
\end{equation}
where $\alpha_{\rm DR}$ [cm$^3$ s$^{-1}$] is the recombination rate coefficient. This recombination rate coefficient includes both the coefficient for two-body recombination,$\alpha_2$ [cm$^3$ s$^{-1}$], 
and three-body recombination, $\alpha_3$ [cm$^6$ s$^{-1}$] such that:
\begin{equation}
\alpha_{\rm DR} = \alpha_2 + n_{\rm gas} \alpha_3.
\label{eqn:recombination-rate}
\end{equation}
The two-body process is taken to be the recombination rate for protonated hydrogen \citep[][their Eq. 7]{McCall2004},
\begin{equation}
\alpha_2 / {\rm \big(cm^3 \, s^{-1}\big)} = 8.22\times10^{-8}\Bigg(\dfrac{T}{300 {\rm K}}\Bigg)^{\!\!-0.48} \!\!\!\!\!\!\!\! - \, 1.3\times10^{-8}.
\label{eqn:alpha2}
\end{equation}
This equation agrees with the measured two body recombination rate for protonated hydrogen over a temperature range of $10$ K $< T < 4000$ K, and is in agreement with the typical two-body dissociative
recombination rates for the upper atmosphere of Earth, according to \citet{Bardsley1968}. The three-body recombination rate is taken to be \citep{Smith1977}:
\begin{equation}
 \alpha_3 / {\rm \big(cm^6 \, s^{-1}\big)} = 2\times10^{-25} \Bigg(\dfrac{T}{300 {\rm K}}\Bigg)^{-2.5}
 \label{eqn:alpha3}
\end{equation}
We calculate the steady-state degree of ionization, $f_{e,CR} = n(e^-)/n_{\rm gas}$, by setting $dn(e^-)/dt = 0$ in Eq. (\ref{eqn:ionization-rate-eqn}), and find:
\begin{equation}
f_{e,CR} = \dfrac{1}{n_{\rm gas}}\sqrt{\dfrac{Q}{\alpha_{\rm DR}}}.
\label{eqn:ionization-fraction-CR}
\end{equation}
We will now combine this degree of ionization with the thermal degree of ionization and explore the resulting abundance of free electrons and their possible effect on the magnetic fields of these example objects.

\section{Results and Discussion}
\label{sec:results}

The total degree of ionization in the atmosphere can now be calculated
by summing the degree of ionization, $f_{e,CR}$ from Eq. \ref{eqn:ionization-fraction-CR}, 
discussed in Sect. \ref{sec:crt}, and the electron fraction 
due to thermal ionization included in the \textsc{Drift-Phoenix} model 
atmosphere. The initial chemical abundances and
the degree of thermal ionization, $f_{e,{\rm thermal}} = n_{e,{\rm
    thermal}}/n_{\rm gas}$, is provided by the \textsc{Drift-Phoenix}
model with $T_{\rm eff} = 1500$ K, $\log g = 3$
(giant gas planet), and $\log g = 5$ (brown dwarf). The total degree of ionization, $f_e$, is then:
\begin{equation}
f_e =f_{e,{\rm thermal}} + f_{e,CR}.
\label{eqn:degree-of-ionizaiton}
\end{equation}
Cosmic rays directly affect the number density of free electrons
(Sect. \ref{sec:e-density}). In
Sect. \ref{sec:b-coupling} we analyze the effect of the free electron
enhancement on the magnetic field coupling by evaluating the magnetic
Reynolds number.

\subsection{Cosmic Ray impact on the Electron Gas Density}
\label{sec:e-density}

\begin{figure}
\centering
\begin{tabular}{cc}
\includegraphics[width=0.43\columnwidth,height=0.6\columnwidth]{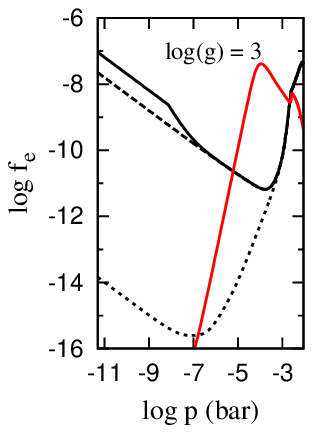} &
\includegraphics[width=0.47\columnwidth,height=0.6\columnwidth]{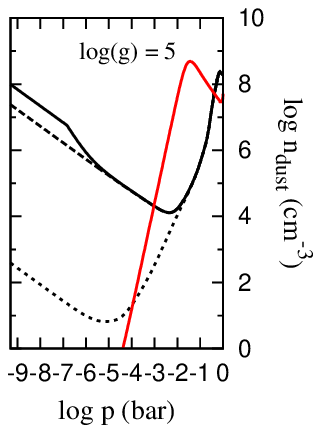}
\end{tabular}
\caption{The degree of gas ionization, $f_e = n(e^-)/n_{\rm gas}$, as a
  function of local gas pressure, for $T_{\rm eff} = 1500$ K, $\log g = 3$
  (giant gas planet, left) and $\log g = 5$ (brown dwarf, right).  The solid line denotes both the
  HECR and LECR contribution, while the dashed line denotes the HECR
  contribution only. The dotted line represents the electron abundance
  in the absence of cosmic rays, and demonstrates the extreme insufficiency of thermal ionization processes in cool objects.
  The solid red line is the dust
  number density, $n_{\rm dust}$ [cm$^{-3}$], and indicates where the cloud layer is located.}
\label{fig:electron-fraction}
\end{figure}

The impact of cosmic rays on the number of free electrons in the
  gas phase is the primary focus of our work.  These results are
plotted in Fig.~\ref{fig:electron-fraction}: For the $\log g = 3$, $T_{\rm eff} = 1500$ K
giant gas planet the HECR component ($10^9$ eV $< E < 10^{12}$ eV) dominates, and cosmic rays
contribute substantially to the ionization until $p_{\rm gas} \sim 10^{-3}$ bar.
For the $\log g = 3$, $T_{\rm eff} = 1000$ K giant gas planet, the HECR
component dominates until $10^{-2}$ bar.
For the $\log g = 5$ case (brown dwarf), the LECR component ($E < 10^9$ eV) dominates
until $p_{\rm gas} \sim 10^{-4}$ bar. The HECR component then takes over until
$p_{\rm gas} \sim 10^{-2}$ bar.  Both the HECR and LECR components are added to
the local degree of thermal ionization that results from the \textsc{Drift-Phoenix} model
atmospheres. The regions where HECR and LECR components dominate for
our three model atmospheres are given in Table~\ref{tab:results}.

Both the HECR and LECR components reach the cloud top, and the HECR component produces the bulk
of free electrons within the upper 10\% of cloud layer for the $\log g = 3$,$T_{\rm eff} = 1500$ K (giant gas planet)
atmosphere, 50\% of cloud layer for the $\log g = 3$, $T_{\rm eff} = 1000$ K (giant gas planet) atmosphere and 5\%
of the cloud layer for the $\log g = 5$, $T_{\rm eff} = 1500$ K (brown dwarf) atmosphere.
In regions where the LECR component dominates, the degree of ionization is enhanced by about a factor of 5 over what
the HECR component would contribute alone. For the $\log g = 3$,$T_{\rm eff} = 1000$ K model atmosphere, the degree
of ionization, $f_e$, exceeds $10^{-8}$ when $p_{\rm gas} < 10^{-8}$ bar. This approaches $f_e \sim 10^{-7}$, when 
the gas begins to act like a weakly ionized plasma \citep{Diver2001}. 
The pressures below which this degree of ionization is reached are
also included in Table~\ref{tab:results}. The brown dwarf atmosphere never achieves such a high degree of ionization.
It is interesting to note in this context that clouds in the Earth atmosphere are not directly ionized by cosmic
rays. Instead, the cosmic rays ionize the gas above the clouds and an ion current develops that leads to the ionization 
of the upper cloud layers \citep{Nicoll2010}. Considering how far cosmic rays penetrate into the cloud layers of our model
atmospheres, it would be useful to explore the charging of grains by the resulting free electrons.

\begin{deluxetable}{lcccccc}
\label{tab:results}
\tablecaption{Effect of Cosmic Ray Ionization on Model Atmospheres by Region (See Fig's \ref{fig:profile}, \ref{fig:electron-fraction} and \ref{fig:electron-fraction-1000})}
\tablenum{1}
\tablehead{ & & $p_{\rm gas}$ at the & $p_{\rm gas}$ at the & $p_{\rm gas}$ where & $p_{\rm gas}$ where & $p_{\rm gas}$ 
where  \\ \colhead{$g$} & \colhead{$T_{\rm eff}$} & \colhead{Cloud Top} & \colhead{Cloud Base} 
& \colhead{HECR$^{\rm a}$} & \colhead{LECR$^{\rm b}$} & \colhead{$f_e > 10^{-8}$} \\ 
\colhead{(cm s$^{-2}$)} & \colhead{(K)} & \colhead{(bar)} & \colhead{(bar)} & \colhead{(bar)} & \colhead{(bar)} & 
\colhead{(bar)} } 
\startdata
3 & 1500 & $5\times10^{-5}$ & $10^{-2}$        &  $10^{-7}-10^{-3}$       & $<10^{-7}$ & $<5\times10^{-10}$ \\
3 & 1000 & $2\times10^{-3}$ & $10^{-1}$        & $10^{-5}-5\times10^{-2}$ & $<10^{-5}$ & $<10^{-8}$ \\
5 & 1500 & $10^{-3}$        & $\sim 1$ & $10^{-6}-3\times10^{-2}$ & $<10^{-6}$ & $--^{\rm c}$ \\
\enddata
\tablecomments{\\$^{\rm a}$The region where cosmic rays of energy $10^9$ eV$ < E < 10^{12}$ eV dominate.\\
$^{\rm b}$The region where cosmic rays of energy $E < 10^9$ eV dominate.\\
$^{\rm c}$ $f_e < 10^{-8}$ throughout this model atmosphere.}
\end{deluxetable}

\begin{figure}
\centering
\includegraphics[width=\columnwidth]{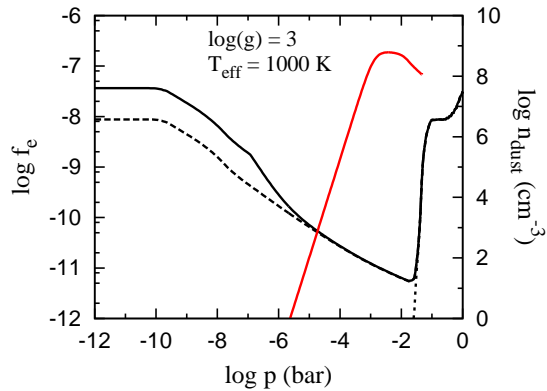}
\caption{The degree of gas ionization, $f_e = n(e^-)/n_{\rm gas}$, as a
  function of local gas pressure, for  $\log g = 3$, $T_{\rm eff} = 1000$ K.  
  The solid line denotes both the
  HECR and LECR contribution, while the dashed line denotes the HECR
  contribution only. The dotted line represents the electron abundance
  in the absence of cosmic rays, hence thermal ionization only.  The solid red line is the dust
  number density, $n_{\rm dust}$ [cm$^{-3}$], and indicates where the cloud layer is located.}
\label{fig:electron-fraction-1000}
\end{figure}

\subsection{Magnetic Field Coupling }
\label{sec:b-coupling}

It is helpful to recast our results in terms of the degree the
magnetic field couples to the gas. \cite{Helling2011} quantify the
degree of coupling by the magnetic Reynolds number, $R_M$, a
dimensionless quantity which is directly proportional to the
atmospheric degree of ionization, $f_e = p_e/p_{\rm gas} = n_e/n_{\rm
  gas}$, where $p_e$ and $p_{\rm gas}$ are the electron pressure and
gas pressure, respectively. The gas pressure and electron pressure are
provided by the \textsc{Drift-Phoenix} model atmospheres where the 
electron pressure is calculated from the Saha Equation for thermal ionization
processes.

\begin{figure}
\centering
\begin{tabular}{cc}
\includegraphics[width=0.43\columnwidth,height=0.6\columnwidth]{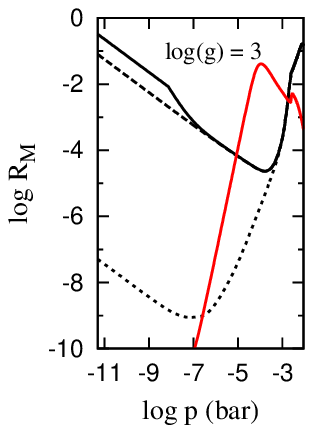} &
\includegraphics[width=0.47\columnwidth,height=0.6\columnwidth]{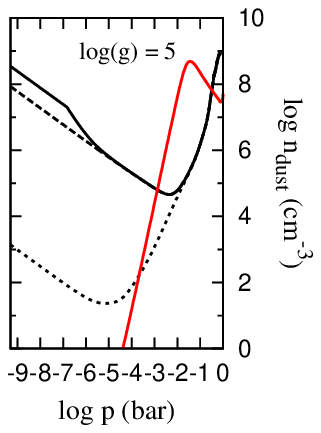}
\end{tabular}
\caption{Magnetic Reynold's number (Eq. \ref{eqn:RM}) for $T_{\rm eff} = 1500$ K, $\log g = 3$ 
(giant gas planet, left) and $\log g = 5$ (brown dwarf, right). 
The solid line denotes both the HECR and LECR contribution, while 
the dashed line denotes the HECR contribution only. The dotted line represents 
the electron abundance in the absence of cosmic rays.
The solid red line is the dust number density, $n_{\rm dust}$ [cm$^{-3}$] 
and indicates where the cloud layer is located.}
\label{fig:magnetic-R}
\end{figure}

The magnetic Reynolds number can be expressed as \citep{Helling2011}:
\begin{equation}
 R_M = (10^{9} \text{ cm$^2$ s$^{-1}$}) \, \dfrac{4\pi q^2}{m_ec^2}\dfrac{1}{\braket{\sigma v}_{\rm en}} \; f_e,
 \label{eqn:RM} 
\end{equation}
where $q$ is the electric charge, $m_e$ is the electron mass, $c$ is
the speed of light, and $\braket{\sigma v}_{\rm en}$ is the
collisional rate, taken to be $\approx 10^{-9}$ cm$^3$ s$^{-1}$. The coupling of
the magnetic field to the gas is expected if $R_M > 1$, and though
cosmic rays enhance $R_M$ in the outer atmosphere by orders of
magnitude, the magnetic Reynolds number does not reach unity anywhere
within the upper atmosphere (Fig. \ref{fig:magnetic-R}). 
Since $R_M$ varies linearly with $f_e$, cosmic rays affect
the magnetic Reynolds number over the same pressure range that they affect the degree of ionization
for all model atmospheres.
Since none of the model atmospheres achieves a value of $R_M > 1$, this suggests that mechanisms 
other than cosmic rays of $E < 10^{12}$ eV will be
needed if the magnetic field is to be coupled to the atmospheric gas in giant gas planets or brown dwarfs.

\begin{figure}
\centering
\includegraphics[width=\columnwidth]{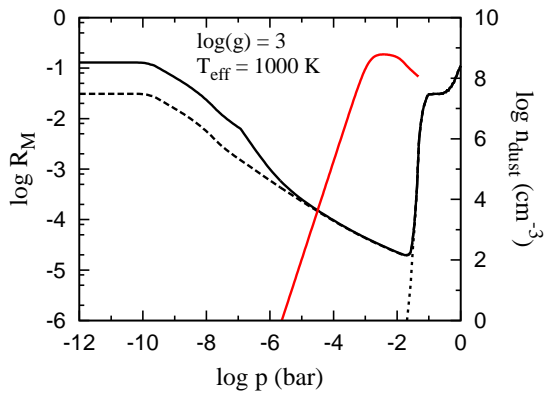}
\caption{Magnetic Reynold's number (Eq. \ref{eqn:RM}) for $\log g = 3$, $T_{\rm eff} = 1000$ K.  
The solid line denotes both the
  HECR and LECR contribution, while the dashed line denotes the HECR
  contribution only. The dotted line represents the electron abundance
  in the absence of cosmic rays, hence thermal ionization only.  The solid red line is the dust
  number density, $n_{\rm dust}$ [cm$^{-3}$], and indicates where the cloud layer is located.}
\label{fig:magnetic-R-1000}
\end{figure}

\section{Summary}

This paper seeks to answer the question of the significance of cosmic
ray ionization on the number of free electrons in brown dwarfs and
giant gas planets. We further examine the possible 
coupling of the magnetic field to
the gas because of the increased degree of ionization.  We develop an analytical
model for the exospheric density profile
which we combine to an atmospheric structure. 
The \textsc{Drift-Phoenix}
model atmospheres provide the inner boundary conditions for the
exosphere model, and they provide the density profile that we utilize
below the exobase.  Cosmic ray transport through the exosphere 
and the atmosphere is
then calculated using these density profiles. A Monte Carlo cosmic ray
transport method from \cite{Rimmer2012} is applied to cosmic rays of $E <
10^9$ eV. An analytic method for cosmic ray transport from \cite{Velinov2008} 
is applied to cosmic rays of $E > 10^9$ eV. We
calculate an ionization rate for which we provide a parameterized 
expression (Eq. \ref{eqn:e-production}). We use this expression to estimate
the steady state degree of ionization (Eq. \ref{eqn:degree-of-ionizaiton}).

Do cosmic rays have a significant impact on the electron fraction? If
the measure of significance is the number of free electrons in the upper atmosphere, then the answer is ``yes''. 
Cosmic ray ionization is responsible for almost all the free electrons in our upper model atmospheres
and for the giant gas planet model atmospheres, achieves a degree of ionization approaching that necessary to qualify the
highest regions of these model atmospheres as weakly interacting plasmas, thereby providing an environment
for plasma processes \citep{Stark2013}. If, however, the measure of significance is the degree of coupling of the
magnetic field to the gas, then the answer is ``no''.  The model
predicts a cosmic ray enhancement to the steady-state electron
abundance by several orders of magnitude 
for atmospheric regions with $p_{\rm gas} < 10^{-3}$ bar for brown dwarf conditions and 
for $p_{\rm gas} < 10^{-2}$,$10^{-4}$ bar for giant gas planet conditions, with $T_{\rm eff} = 1500$ K in both cases.
This enhancement is not large enough to allow the magnetic Reynolds
number, $R_M >1$ above the cloud top, and does not
significantly affect $R_M$ for the bulk of the cloud layer. This
indicates that the magnetic field would not couple to the gas because
of the steady-state cosmic ray ionization enhancement.  However,
  the geometry of the magnetic field \citep[e.g.][]{Donati2008,Vidotto2012,Lang2012} 
 might lead to a
  channeling of the cosmic rays which then would amplify the local
  degree of ionization beyond the values determined in this paper,
which would cause heterogeneous distribution of cosmic ray induced
  chemical products. The increased abundance of electrons may contribute to charge build-up
on dust at the top of the cloud layer. This is especially the case for
our model brown dwarf atmosphere, because cosmic rays penetrate more
deeply into its cloud layer.

Cosmic rays were first discovered by their 
ionization effects in the Earth's atmosphere. We
predict that ionization effects have a significant impact on the
upper atmospheres of free-floating extrasolar planets 
and very low-mass stars. The effect the cosmic ray induced degree of ionization has on the chemistry
in the upper atmospheres of these objects is a very interesting question. It has been suggested that
cosmic rays may drive production of small hydrocarbons that may be responsible for the hazes of e.g. HD 189733 b 
\citep{Moses2011}. This is a question we hope to explore in detail in a future paper.

\acknowledgments
We highlight financial support of the European Community under the FP7 by an ERC starting grant.
We are grateful to Kristina Kislyakova and Helmut Lammer for helpful discussions relating to our exosphere model.




\bibliographystyle{apj}

\end{document}